\documentclass[twocolumn,twocolappendix,times,astrosymb,tighten]{aastex701}

\usepackage{hyperref}
\usepackage{CJKutf8}
\usepackage{amsmath}
\hypersetup{linkcolor=red,citecolor=blue,filecolor=cyan,urlcolor=blue}

\newcommand{\zh}[1]{\begin{CJK}{UTF8}{gbsn}#1\end{CJK}}

\newcommand{\frb}{FRB\,20250316A}

\newcommand{\be}{\begin{equation}}
\newcommand{\ee}{\end{equation}}

\graphicspath{{./}{figures/}}

\shorttitle{JWST Observations of \frb}
\shortauthors{Blanchard et al.}

\begin{document}

\title{\textit{\textbf{James Webb Space Telescope}} Observations of the Nearby and Precisely-Localized \frb: A Potential Near-IR Counterpart and Implications for the Progenitors of Fast Radio Bursts}

\correspondingauthor{Peter K.~Blanchard}
\email{peter.blanchard@cfa.harvard.edu}

\author[0000-0003-0526-2248]{Peter K.~Blanchard}
\email{peter.blanchard@cfa.harvard.edu}
\affil{Center for Astrophysics \(|\) Harvard \& Smithsonian, 60 Garden St., Cambridge, MA 02138, USA}

\author[0000-0002-9392-9681]{Edo Berger}
\email{eberger@cfa.harvard.edu}
\affil{Center for Astrophysics \(|\) Harvard \& Smithsonian, 60 Garden St., Cambridge, MA 02138, USA}

\author[0000-0002-3980-815X]{Shion E.~Andrew}
\email{shiona@mit.edu}
  \affiliation{MIT Kavli Institute for Astrophysics and Space Research, Massachusetts Institute of Technology, 77 Massachusetts Ave, Cambridge, MA 02139, USA}
  \affiliation{Department of Physics, Massachusetts Institute of Technology, 77 Massachusetts Ave, Cambridge, MA 02139, USA}

\author[0009-0005-8230-030X]{Aswin Suresh}
\email{aswinsuresh2029@u.northwestern.edu}
\affiliation{Department of Physics and Astronomy, Northwestern University, Evanston, IL 60208, USA}
\affiliation{Center for Interdisciplinary Exploration and Research in Astrophysics, Northwestern University, 1800 Sherman Avenue, Evanston, IL 60201, USA}

\author[0000-0002-6765-8988]{Kohki Uno}
\email{k.uno@kusastro.kyoto-u.ac.jp}
\affil{Department of Astronomy, Kyoto University, Kitashirakawa-Oiwake-cho, Sakyo-ku, Kyoto, 606-8502, Japan}

\author[0000-0002-5740-7747]{Charles D.~Kilpatrick}
\email{ckilpatrick@northwestern.edu}
\affiliation{Center for Interdisciplinary Exploration and Research in Astrophysics, Northwestern University, 1800 Sherman Avenue, Evanston, IL 60201, USA}

\author[0000-0002-4670-7509]{Brian D.~Metzger}
\email{bdm2129@columbia.edu}
\affil{Department of Physics and Columbia Astrophysics Laboratory, Columbia University, Pupin Hall, New York, NY 10027, USA}
\affil{Center for Computational Astrophysics, Flatiron Institute, 162 5th Ave, New York, NY 10010, USA}

\author[0000-0003-0871-4641]{Harsh Kumar}
\email{harsh.kumar@cfa.harvard.edu}
\affil{Center for Astrophysics \(|\) Harvard \& Smithsonian, 60 Garden St., Cambridge, MA 02138, USA}

\author[0000-0002-5519-9550]{Navin Sridhar}
\email{NAVINSRIDHAR@stanford.edu}
\affil{Department of Physics, Stanford University, 382 Via Pueblo Mall, Stanford, CA 94305, USA}
\affil{Kavli Institute for Particle Astrophysics \& Cosmology, P.O. Box 2450, Stanford University, Stanford, CA 94305, USA}

\author[0000-0001-6422-8125]{Amanda M.~Cook}
  \email{amanda.cook@mail.mcgill.ca}
  \affiliation{Department of Physics, McGill University, 3600 rue University, Montr\'eal, QC H3A 2T8, Canada}
  \affiliation{Trottier Space Institute, McGill University, 3550 rue University, Montr\'eal, QC H3A 2A7, Canada}
  \affiliation{Anton Pannekoek Institute for Astronomy, University of Amsterdam, Science Park 904, 1098 XH Amsterdam, The Netherlands}

\author[0000-0002-9363-8606]{Yuxin (\zh{董雨欣}) Dong}
  \email{yuxin.dong@northwestern.edu}
  \affiliation{Department of Physics and Astronomy, Northwestern University, Evanston, IL 60208, USA}
  \affiliation{Center for Interdisciplinary Exploration and Research in Astrophysics, Northwestern University, 1800 Sherman Avenue, Evanston, IL 60201, USA}
  
\author[0000-0003-0307-9984]{Tarraneh Eftekhari}
  \email{teftekhari@northwestern.edu}
  \affiliation{Center for Interdisciplinary Exploration and Research in Astrophysics, Northwestern University, 1800 Sherman Avenue, Evanston, IL 60201, USA}
  
\author[0000-0002-7374-935X]{Wen-fai Fong}
  \email{wfong@northwestern.edu}
  \affiliation{Department of Physics and Astronomy, Northwestern University, Evanston, IL 60208, USA}
  \affiliation{Center for Interdisciplinary Exploration and Research in Astrophysics, Northwestern University, 1800 Sherman Avenue, Evanston, IL 60201, USA}

\author[0000-0001-7946-1034]{Walter W.~Golay}
\email{walter.golay@cfa.harvard.edu}
\affil{Center for Astrophysics \(|\) Harvard \& Smithsonian, 60 Garden St., Cambridge, MA 02138, USA}

\author[0000-0002-1125-9187]{Daichi Hiramatsu}
\email{daichi.hiramatsu@cfa.harvard.edu}
\affil{Center for Astrophysics \(|\) Harvard \& Smithsonian, 60 Garden St., Cambridge, MA 02138, USA}

\author[0000-0003-3457-4670]{Ronniy C.~Joseph}
  \email{ronniy.joseph@mcgill.ca}
  \affiliation{Department of Physics, McGill University, 3600 rue University, Montr\'eal, QC H3A 2T8, Canada}
  \affiliation{Trottier Space Institute, McGill University, 3550 rue University, Montr\'eal, QC H3A 2A7, Canada}

\author[0000-0001-9345-0307]{Victoria M.~Kaspi}
  \email{victoria.kaspi@mcgill.ca}
  \affiliation{Department of Physics, McGill University, 3600 rue University, Montr\'eal, QC H3A 2T8, Canada}
  \affiliation{Trottier Space Institute, McGill University, 3550 rue University, Montr\'eal, QC H3A 2A7, Canada}

\author[0000-0002-5857-4264]{Mattias Lazda}
  \email{mattias.lazda@mail.utoronto.ca}
  \affiliation{Dunlap Institute for Astronomy and Astrophysics, 50 St. George Street, University of Toronto, ON M5S 3H4, Canada}
  \affiliation{David A.\ Dunlap Department of Astronomy and Astrophysics, 50 St. George Street, University of Toronto, ON M5S 3H4, Canada}

\author[0000-0002-4209-7408]{Calvin Leung}
  \email{calvin_leung@berkeley.edu}
  \affiliation{Miller Institute for Basic Research, Stanley Hall, Room 206B, Berkeley, CA 94720, USA}
  \affiliation{Department of Astronomy, University of California, Berkeley, CA 94720, USA}

\author[0000-0002-4279-6946]{Kiyoshi W.~Masui}
  \email{kmasui@mit.edu}
  \affiliation{MIT Kavli Institute for Astrophysics and Space Research, Massachusetts Institute of Technology, 77 Massachusetts Ave, Cambridge, MA 02139, USA}
  \affiliation{Department of Physics, Massachusetts Institute of Technology, 77 Massachusetts Ave, Cambridge, MA 02139, USA}

\author[0000-0002-0772-9326]{Juan Mena-Parra}
  \email{juan.menaparra@utoronto.ca}
  \affiliation{Dunlap Institute for Astronomy and Astrophysics, 50 St. George Street, University of Toronto, ON M5S 3H4, Canada}
  \affiliation{David A.\ Dunlap Department of Astronomy and Astrophysics, 50 St. George Street, University of Toronto, ON M5S 3H4, Canada}

\author[0000-0003-0510-0740]{Kenzie Nimmo}
  \email{knimmo@mit.edu}
  \affiliation{MIT Kavli Institute for Astrophysics and Space Research, Massachusetts Institute of Technology, 77 Massachusetts Ave, Cambridge, MA 02139, USA}

\author[0000-0002-8912-0732]{Aaron B.~Pearlman}
  \email{aaron.b.pearlman@physics.mcgill.ca}
  \affiliation{Department of Physics, McGill University, 3600 rue University, Montr\'eal, QC H3A 2T8, Canada}
  \affiliation{Trottier Space Institute, McGill University, 3550 rue University, Montr\'eal, QC H3A 2A7, Canada}
  \altaffiliation{Banting Fellow, McGill Space Institute Fellow, and FRQNT Postdoctoral Fellow}

\author[0000-0002-4823-1946]{Vishwangi Shah}
  \email{vishwangi.shah@mail.mcgill.ca}
  \affiliation{Department of Physics, McGill University, 3600 rue University, Montr\'eal, QC H3A 2T8, Canada}
  \affiliation{Trottier Space Institute, McGill University, 3550 rue University, Montr\'eal, QC H3A 2A7, Canada}
  
\author[0000-0002-6823-2073]{Kaitlyn Shin}
  \email{kshin@mit.edu}
  \affiliation{MIT Kavli Institute for Astrophysics and Space Research, Massachusetts Institute of Technology, 77 Massachusetts Ave, Cambridge, MA 02139, USA}
  \affiliation{Department of Physics, Massachusetts Institute of Technology, 77 Massachusetts Ave, Cambridge, MA 02139, USA}

\author[0000-0003-3801-1496]{Sunil Simha}
  \email{sunil.simha@northwestern.edu}
  \affiliation{Center for Interdisciplinary Exploration and Research in Astrophysics, Northwestern University, 1800 Sherman Avenue, Evanston, IL 60201, USA}
  \affiliation{Department of Astronomy and Astrophysics, University of Chicago, William Eckhardt Research Center, 5640 S Ellis Ave, Chicago, IL 60637, USA}

\begin{abstract}
    We present deep {\it James Webb Space Telescope} near-infrared imaging to search for a quiescent or transient counterpart to \frb, which was precisely localized with the CHIME/FRB Outriggers array to an area of $11\times13$ pc in the outer regions of NGC\,4141 at $d\approx40$ Mpc.  Our F150W2 image reveals a faint source near the center of the FRB localization region (``NIR-1''; $M_{\rm F150W2}\approx-2.5$ mag; probability of chance coincidence $\approx0.36$), the only source within $\approx2.7\sigma$. We find that it is too faint to be a globular cluster, young star cluster, red supergiant star, or a giant star near the tip of the red giant branch (RGB).  It is instead consistent with a red giant near the RGB ``clump'' or a massive ($\gtrsim20$ M$_{\odot}$) main sequence star, although the latter explanation is less likely. The source is too bright to be a supernova remnant, Crab-like pulsar wind nebula, or isolated magnetar. Alternatively, NIR-1 may represent transient emission, namely a dust echo from an energetic outburst associated with the FRB, in which case we would expect it to fade in future observations.   We explore the stellar population near the FRB and find that it is composed of a mix of young massive stars ($\sim10-100$ Myr) in a nearby \ion{H}{2} region that extends to the location of \frb, and old evolved stars ($\gtrsim$ Gyr). The overlap with a young stellar population, containing stars of up to $\approx20$ M$_\odot$, may implicate a neutron star / magnetar produced in the core collapse of a massive star as the source of \frb.
\end{abstract}


\section{Introduction}

Fast radio bursts (FRBs) are bright, millisecond-duration pulses at $\sim$GHz frequencies, whose origin remains a mystery \citep{Lorimer2007Sci...318..777L,Petroff2019A&ARv..27....4P,Petroff2022A&ARv..30....2P}. Their dispersion measures (DM), the integrated electron column density along the line of sight, significantly exceed the range of the Milky Way and its halo, implying an extragalactic origin, which has now been confirmed with localizations of over 100 FRBs to galaxies at a wide range of redshifts \citep{Chatterjee2017Natur.541...58C,Heintz2020ApJ...903..152H,Bhandari2022AJ....163...69B,Gordon2023,Bhardwaj2024,Law2024ApJ...967...29L,Sharma2024,Gordon2025}. While most FRBs appear as one-off events, a growing subset are known to repeat \citep{Spitler2016Natur.531..202S,CHIME/FRB2019ApJ...885L..24C,Fonseca2020ApJ...891L...6F,CHIME2023ApJ...947...83C}, and it is possible that the entire population repeats on a wide range of timescales (e.g., \citealt{Ravi2019NatAs...3..928R,James2023}).

Despite the rapidly increasing sample and the availability of localizations and host galaxy properties, the physical origin(s) and production mechanism(s) of FRBs remain elusive, with dozens of proposed sources and models suggested to date (e.g., \citealt{Platts2019PhR...821....1P}). Given the short duration (and non-catastrophic nature of the repeating FRBs) models involving neutron stars or black holes are particularly popular; FRB-like detections from the Galactic magnetar SGR~1935+2154 support such a scenario \citep{CHIME/FRB2020Natur.587...54C,Bochenek2020Natur.587...59B}. The mix of host galaxy types, and the fact that some FRBs show no correlation with star formation (e.g., 
\citealt{Kirsten2022Natur.602..585K,Eftekhari2025,Shah2025ApJ...979L..21S}), is suggestive of compact objects formed through both young and delayed channels (e.g., \citealt{Margalit2019ApJ...886..110M,Sridhar+21,Kremer2023ApJ...944....6K,Horowicz2025}).

A significant barrier to deciphering the origin of FRBs is their sole detection in the radio band. Temporally coincident X-ray bursts from the FRB-like Galactic SGR~1935+2154 remain the only non-radio transient detections \citep{Mereghetti2020ApJ...898L..29M,Ridnaia2021NatAs...5..372R,Li2021NatAs...5..378L,Tavani2021NatAs...5..401T}, but such bursts would not be detectable at the distances of extragalactic FRBs  \citep{Scholz2016ApJ...833..177S,Scholz2017ApJ...846...80S,Scholz2020ApJ...901..165S,Tavani2020ApJ...893L..42T,Cook2024,Pearlman2025}.  Searches for prompt FRB counterparts at non-radio wavelengths have been carried out, but to date no detections have been made \citep{Hardy2017MNRAS.472.2800H,MAGIC2018MNRAS.481.2479M,Kilpatrick2021ApJ...907L...3K,Lin2020Natur.587...63L,Niino2022ApJ...931..109N,Hiramatsu2023ApJ...947L..28H,Kilpatrick2024ApJ...964..121K,Curtin2023,Curtin2024}. 

An alternative approach is to search for ``quiescent'' emission from the progenitor system itself, specifically from a stellar companion.  A search for such lower-luminosity quiescent counterparts requires a precise localization and a nearby FRB, in order to place direct meaningful limits on a progenitor system. This was the case for FRB\,20200120E, precisely localized to a globular cluster in M\,81, but due to the high stellar density in the cluster a unique counterpart could not be identified \citep{Dage2023}.  Persistent radio sources have been identified for a small subset of {\it repeating} FRBs (e.g., \citealt{Chatterjee2017Natur.541...58C,Niu2022Natur.606..873N}), but their origin and relation to the FRB-emitting sources remain unclear.

Against this backdrop, on 2025 March 16, CHIME/FRB detected the bright \frb, and subsequently localized it using the CHIME/FRB Outriggers array to an area of only $\approx 0.012$ arcsec$^2$ ($\approx 440$ pc$^{2}$) in the outskirts of the galaxy NGC\,4141 at $d\approx 40$ Mpc \citep{Cook25}.  \frb\ therefore presents a unique opportunity to explore the progenitor system of an FRB, or any associated delayed transient emission. Initial observations in the optical and X-rays, as well as archival radio data, did not reveal a counterpart \citep{Cook25,mmt,Golay_Atel_2025,chandra}.  

Here, we present a deep search for an infrared (IR) counterpart using the {\it James Webb Space Telescope} ({\it JWST}), which significantly extends the reach of counterpart searches to \frb. The paper is organized as follows.  In \S\ref{sec:obs} we describe the {\it JWST} observations, astrometric alignment, photometry, and identification of a potential counterpart.  In \S\ref{sec:comps} we assess possible sources of quiescent and transient NIR emission, as well as the properties of the surrounding stellar population. We summarize the key findings in \S\ref{sec:conc}. Throughout the paper we use AB magnitudes and $d=40$ Mpc ($1''=190$ pc) for the distance to NGC\,4141.  There is negligible Galactic extinction in the {\it JWST} bands in the direction of \frb\ ($A_{\rm F150W2}\approx 0.01$ mag).

\section{{\it JWST} Observations}
\label{sec:obs}

Following the CHIME/FRB Outrigger localization of \frb, we were awarded Director's Discretionary Time to image the FRB localization region with {\it JWST} (Program ID: 9331, PI: Blanchard). We used the NIRCam instrument starting on 2025 May 19 at 13:55 UT to obtain a 9-point standard sub-pixel dither pattern for a total on-source science exposure time of 9470 s, with the extra-wide filters F150W2 ($1.0-2.4$ $\mu$m) and F322W2 ($2.4-4.0$ $\mu$m). We retrieved the 9 calibrated dithered (level-2) images for each filter from the Mikulski Archive for Space Telescopes\footnote{https://archive.stsci.edu}.  

Using the {\it JWST} Science Calibration Pipeline we generate two drizzle-combined (level-3) images in each filter, one drizzled to the native pixel scale ($\approx 30$ and $\approx 60$ mas pixel$^{-1}$ for F150W2 and F322W2, respectively) and one drizzled to a finer grid of $20$ and $40$ mas pixel$^{-1}$, respectively (using {\tt pixfrac} $= 0.9$).  The level-2 frames are aligned relative to each other using {\tt JHAT} \citep{JHAT} before drizzling to the level-3 images. 

\begin{figure*}[t!]
    \centering
    \includegraphics[width=0.97\linewidth]{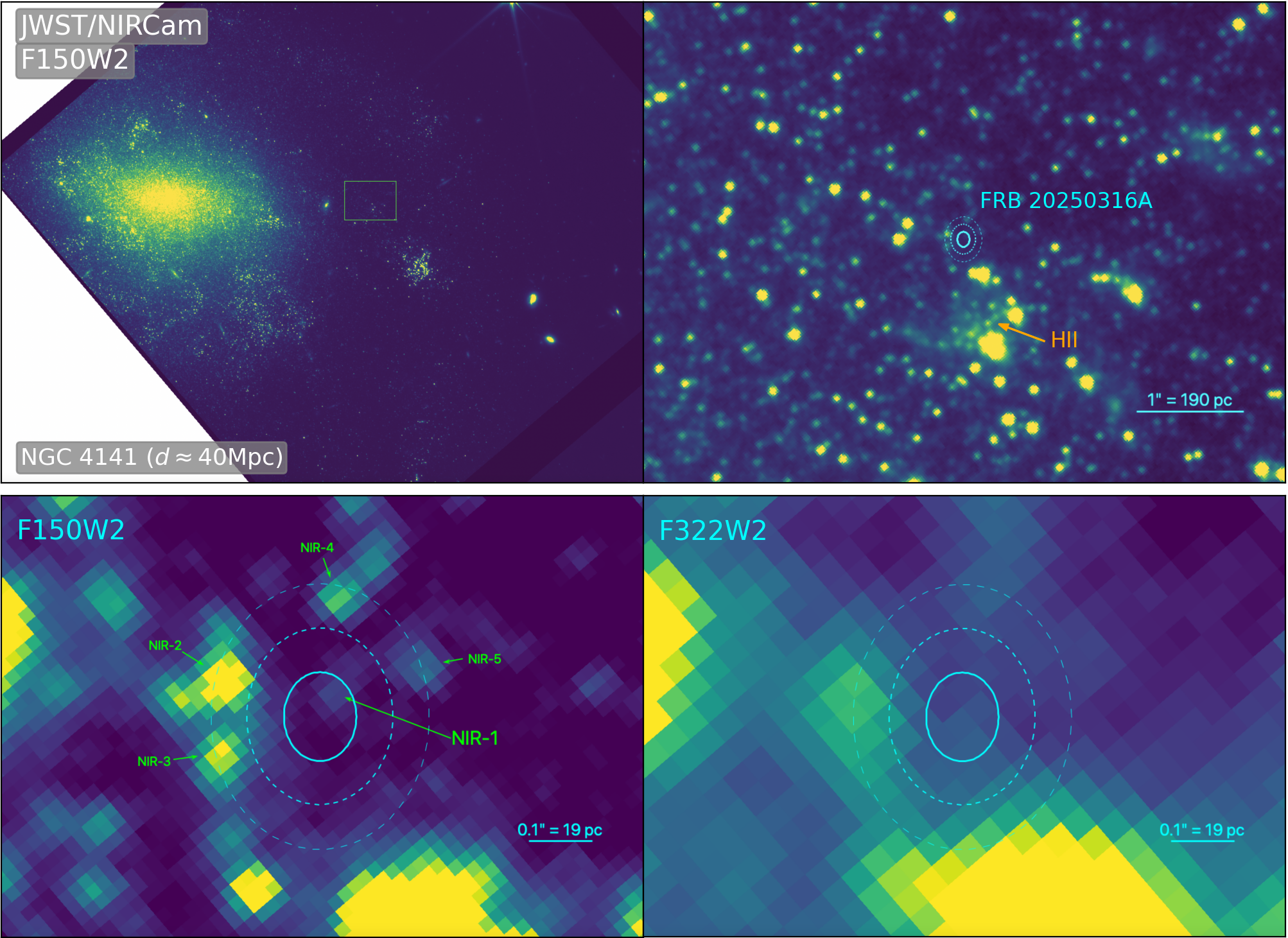}
    \caption{{\it Top left:} Our {\it JWST}/NIRCam F150W2 image of NGC\,4141, the host galaxy of \frb, zoomed to highlight the large-scale structure of the galaxy.  {\it Top right:} Zoomed-in region corresponding to the box in the left panel showing the $1,2,3\sigma$ localization regions of \frb\ in relation to surrounding resolved stellar population, including a nearby \ion{H}{2} region, which is resolved into individual stars and a faint diffuse background.  {\it Bottom:} Further zoom-in on the location of \frb, revealing a single faint source, NIR-1, near the center of the FRB localization region in the F150W2 image ({\it Left}); the only other detected sources are on the edge of the $3\sigma$ region (marked NIR-2 to NIR-5).  NIR-1 is not clearly identified in the lower-resolution F322W2 image ({\it Right}).  The images are aligned with north up and east to the left.}
    \label{fig:images}
\end{figure*}

\subsection{Astrometry}

To precisely match our NIRCam images to the same astrometric ICRF frame as the CHIME/FRB observations \citep[see][for a description of their astrometric calibration]{Cook25}, we aligned the level-3 F150W2 and F322W2 images in their native pixel scale to {\it Gaia} astrometric calibrators.  There are four isolated, point-like\footnote{{\it Gaia} point source scores of {\tt PSS} $>0.99$.  An additional {\it Gaia} source (ID 1575927168528652800) is in both {\it JWST} images but has {\tt PSS} $=0.951$ and is resolved into at least four distinct sources in the {\it JWST} images, and so we do not include it in the procedure below.} {\it Gaia} astrometric calibrators\footnote{{\it Gaia} DR3 source IDs 1575927065449950848, 1575927095514978304, 1575927168529272320, and 1575927237248643840.}   across both the F150W2 and F322W2 images.  When performing {\tt dolphot} photometry (\S\ref{sec:photometry}), we confirm that all four are point-like sources in the {\it JWST} images ({\tt dolphot type=1}).  In particular, one of these sources is the core of NGC\,4141, which appears unresolved and point-like in both {\it Gaia} and {\it JWST}.  Two of our four astrometric calibrators are unsaturated in the {\tt dolphot} point-spread function (PSF) photometry, and so we use the centroids derived from this procedure. The other two are saturated in both images, and we determine their centroids using the unsaturated pixels in the level-2 F150W2 frames, based on the instrumental NIRCam PSF available in {\tt STPSF} \citep{STPSF}. Using this instrumental PSF with the {\tt photutils.psf} package \citep{photutils}, we derive a centroid and uncertainty for both stars, with an uncertainty of $\approx 1.5$ mas.

We consider the following sources of astrometric uncertainty when aligning the F150W2 and F322W2 images to the {\it Gaia} astrometric frame: the statistical uncertainty from the fit itself, the systematic uncertainty associated with our fitting method, and the uncertainty in the relative alignment between all NIRCam images.  To estimate these sources of uncertainty, we derive the astrometric transformation from the level-3 F150W2 instrumental frame to the {\it Gaia} frame using two methods: (i) we fix the image pixel scale and position angle of the image frame to the value reported by on-board telemetry of {\it JWST} and derive the x,y shifts using the four {\it Gaia} standards weighted by the uncertainty in our derived centroids; and (ii) performing the same procedure, but with the image rotation as a free parameter.  For method (i) we find a shift of $-18.0$ mas in R.A.~and $-34.6$ mas in Decl., with associated dispersion of 6.7 mas in R.A.~and 15.3 mas in Decl., derived by minimizing the root-mean square offset between the derived centroids and {\it Gaia} catalog centroids. We take the dispersion to be the statistical uncertainty on the fit. The difference between the location of the FRB in methods (i) and (ii) is small --- 3.5 mas in R.A.~and 1.2 mas in Decl.~--- and we add this to our astrometric error budget as the systematic uncertainty from our fitting method. Finally, the relative alignment of the level-3 F322W2 image to the F150W2 image in the native pixel scale is known to a precision of 10 mas (i.e., only one-sixth of a pixel in the F322W2 image); we note that this has no impact on the astrometry of the F150W2 image.

Adding the statistical and systematic astrometric uncertainties in quadrature, the total astrometric tie uncertainty is 7.5 mas in R.A.~and 15.3 mas in Decl.  Combining this with the FRB localization region centered at R.A.~$=12\rlap{}^{\rm h}09\rlap{}^{\rm m}44\rlap{.}^{\rm s}319$ and Decl.~$=+58\rlap{}^{\rm \circ}50\rlap{}'56\rlap{.}''708$ with a localization uncertainty of $57\times 68$ mas \citep[oriented $-0.26^{\circ}$ east of north, see][]{Cook25}, we infer a NIRCam-frame localization uncertainty of $58\times 70$ mas, or $11.0\times 13.3$ pc (radius, $1\sigma$).  In Figure \ref{fig:images} we show this region (along with $2\sigma$ and $3\sigma$ regions), on the {\it Gaia}-aligned {\it JWST} images. 

\subsection{Photometry}
\label{sec:photometry}

To obtain precise photometry of sources in the crowded field near \frb, we use {\tt dolphot} \citep{dolphot} on the individual level-2 frames for source detection and PSF photometry using the {\tt WebbPSF} v1.2.1 PSF model \citep{STPSF}.  We use the aligned level-3 F150W2 image drizzled to the native pixel scale of NIRCam as a reference to define the locations of individual sources in a common coordinate system; {\tt dolphot} performs all source detection and photometry in the level-2 images where the PSF shape and flux scale are preserved prior to resampling and drizzling.  These methods have been shown to accurately recover point sources in crowded fields at signal-to-noise thresholds close to or better than predictions from the pre-launch {\it JWST} exposure time calculator \citep[see description in][]{JWST-dolphot}.  The final combined photometry of the detected sources is the inverse variance weighted average of the fluxes from each individual image.

To obtain a precise alignment solution between the level-2 NIRCam images and the level-3 reference image, we use {\tt dolphot}'s frame-to-frame alignment using {\tt Align} $=4$ (shift and distortion with a third-order polynomial), resulting in a typical alignment uncertainty of 1.0\,mas for the F150W2 images and 2.3\,mas for the F322W2 images.  We then find sources in the NIRCam images to a threshold of $2.5\sigma$ and perform PSF photometry with a local background measurement inside each PSF ({\tt FitSky} $=2$). The final photometry catalog is derived using a threshold of $3.5\sigma$ and is further trimmed of spurious sources with {\tt sharpness}$^2 >0.1$, {\tt crowding} $>2$ mag, and quality flags indicating that a source is saturated or narrower than the PSF ({\tt flag} $>3$). Finally, we derive the $5\sigma$ limiting magnitude in the NIRCam images within a $3''$ radius of the FRB location using the average magnitude of sources detected at $5\sigma$ significance.  We find $m_{\rm F150W2}\approx 31.0$ mag and $m_{\rm F322W2}\approx 30.4$ mag.

\subsection{Identification of a Potential Counterpart}
\label{sec:id}

We identify a single faint source in the F150W2 image within the $1\sigma$ and $2\sigma$ localization regions of \frb, located $\approx 40$ mas from the centroid of the FRB location (i.e., within the $1\sigma$ region); we refer to this source as ``NIR-1'' (Figure~\ref{fig:images}).  We measure the following magnitudes for NIR-1: $m_{\rm F150W2} = 30.52\pm 0.14$ mag and $m_{\rm F322W2} = 31.10 \pm 0.40$ mag.  We note that the source detected by {\tt dolphot} in the F322W2 filter at the position of NIR-1 is sub-threshold; we therefore use the $3\sigma$ upper limit of $m_{\rm F322W2}\approx 30.9$ for the subsequent analysis.  At the distance of NGC\,4141, the F150W2 detection corresponds to an absolute magnitude of $M_{\rm F150W2} = -2.48\pm 0.14$ mag.

If we expand the localization region to $3\sigma$, we find four additional sources, all on the edge of the $3\sigma$ region, and hence less likely to be associated with \frb.  We therefore consider NIR-1 as a more likely counterpart compared to these four sources, although we note in the next section any comparisons with possible progenitor scenarios that may be affected.  These sources are somewhat brighter than NIR-1, with magnitudes of $m_{\rm F150W2} = 27.85 \pm 0.01$ (``NIR-2''), $28.69 \pm 0.03$ (``NIR-3''), $28.86 \pm 0.03$ (``NIR-4''), $29.61 \pm 0.06$ (``NIR-5''); and $m_{\rm F322W2} = 29.21 \pm 0.07$, $29.51 \pm 0.10$, $29.97 \pm 0.14$, $30.90 \pm 0.33$.  The brightest of these sources, NIR-2, appears blended with two fainter sources, although it is unclear if these are separate sources or part of an extended underlying structure. 

While NIR-1 is the most likely counterpart of the detected sources, the crowded nature of the field means that there is a non-negligible probability that the FRB localization region coincides with NIR-1 by chance.  We use the {\tt dolphot}-identified sources to calculate the number density of sources at $\lesssim 30.5$ mag within a radius of $3''$ of \frb\ to empirically establish the chance-coincidence probability. We find a number density of $\approx 37$ arcsec$^{-2}$, indicating a chance coincidence probability of $P_{\rm cc}=1-e^{-N} \approx 0.36$, where $N\approx37$ arcsec$^{-2}\times0.012$ arcsec$^{2}\approx0.45$ is the expected number of sources in the $1\sigma$ localization region.  The relatively large value of $P_{\rm cc}$ might be indicative of a spurious alignment, and that the actual counterpart of \frb\ is below the detection limit of our observation. We therefore use the brightness of NIR-1 to both investigate its origin, and to place limits on a NIR counterpart to \frb. Since sources NIR-2 to NIR-5 are all located on the outskirts of the $3\sigma$ localization region, we do not consider them likely direct counterparts of \frb, but we consider them in relation to the overall stellar population in the environment of \frb.  

\section{Potential Sources of NIR Emission}
\label{sec:comps}

We consider the nature of NIR-1, and the resulting implications for \frb, by comparing its luminosity --- $M_{\rm F150W2}\approx -2.5$ mag, $L_\nu\approx 4.4\times 10^{21}$ erg s$^{-1}$ Hz$^{-1}$ ---  and color --- $m_{\rm F150W2}-m_{\rm F322W2}\lesssim -0.4$ mag --- with potential sources of emission.  We first explore ``quiescent'' counterparts, including globular and star clusters, stellar binary companions, a supernova remnant or pulsar wind nebula, and an isolated magnetar (\S\ref{sec:GC}--\ref{sec:magnetar}).  We then investigate the properties of the surrounding stellar population as a way of shedding light on the FRB-producing compact object in the scenario where NIR-1 is a chance coincidence (\S\ref{sec:pop}). Finally, we assess the possibility that the emission is transient in nature, due to an afterglow or a dust echo associated with an energetic outflow produced during the FRB (\S\ref{sec:ag} and \ref{sec:dust}).

\subsection{Globular Cluster}
\label{sec:GC}

Motivated by the precise localization of FRB\,20200120E to a globular cluster (GC) in M\,81 \citep{Bhardwaj2021ApJ...910L..18B,Kirsten2022Natur.602..585K}, we assess a globular cluster origin for the faint NIR source.  In Figure~\ref{fig:dists}, we show the $K$-band luminosity function of GCs in M\,31, which spans $\approx -12$ to $-6$ mag, with a peak at $\approx -8.5$ mag (similar to Milky Way GCs; \citealt{Peacock2010}). Even the faintest GCs are $\gtrsim 25$ times brighter than NIR-1.  Additionally, the GC associated with FRB\,20200120E is $\sim 7$ mag brighter than NIR-1. The nearest sources to the location of \frb\ that reach the brightness of typical GCs are $\approx 0.4''$ or $\approx 80$ pc away, much larger than the typical size of GCs (few pc), indicating that a coincident (or offset) GC origin is unlikely for \frb.

\subsection{Young Star Clusters}
\label{sec:cluster}

\begin{figure}[t!]
    \centering
    \includegraphics[width=1.0\linewidth]{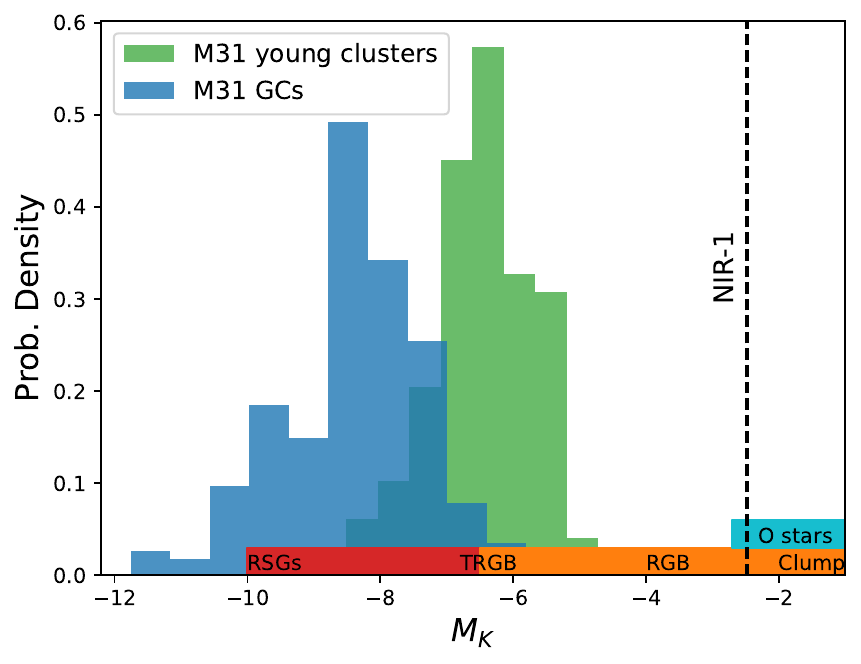}
    \caption{$K$-band luminosity distributions of globular clusters (blue) and young clusters (green) in M31, as well as the approximate luminosity ranges of red supergiants (RSGs; red), stars on the red giant branch (RGB; orange), including the RGB ``tip'' and ``clump'', and main sequence O stars (cyan).  The F150W2 absolute magnitude of NIR-1 (dashed vertical line) indicates it is not a stellar cluster or an RSG, and is instead consistent with the luminosities of red giants near the clump or the massive end of main-sequence O stars.}
    \label{fig:dists}
\end{figure}

We also explore the possibility that NIR-1 is a young, potentially dust-embedded star cluster.  In Figure~\ref{fig:dists} we show the $K$-band luminosity distribution of young clusters in M31, which span $\approx -8.5$ to $-5$ mag \citep{Peacock2010}.  Recent {\it JWST} observations of nearby galaxies have further explored the NIR luminosity functions of young clusters. For example, \citet{Whitmore2023} analyzed 37 such clusters in NGC\,1365, many identified with {\it JWST} for the first time, selected based on their colors and concentration.  These clusters span absolute magnitudes of $\approx -14$ to $-7$ in F200W and $\approx -13$ to $-4$ mag in F335W, all significantly brighter than NIR-1. The {\it JWST} luminosity function inferred from a larger sample of clusters in NGC\,7496 extends slightly fainter to $\approx-4.5$ mag in F200W and $\approx-2.5$ mag in F300W, corresponding to masses down to $\approx 10^{3}$ M$_{\odot}$ \citep{Rodriguez2023}.  NIR-1 is fainter in both F150W2 and F322W2 than the faintest star clusters identified in these galaxies.  In addition, the lack of bright emission in F322W2, which covers the $3.3$ $\mu$m PAH emission, rules out an embedded cluster.  Thus, a star cluster origin is unlikely for \frb.

\subsection{Main Sequence and Evolved Binary Companions}
\label{sec:binary}

Our images are sufficiently deep to place direct constraints on a stellar companion to the FRB-producing compact object (Figure~\ref{fig:dists}). At the brightness of NIR-1, and a magnitude limit of $\approx -2$ mag, we can rule out red supergiant (RSG) stars, which have $K$-band absolute magnitudes in the range of $\approx -10$ to $-6.5$ \citep{Neugent2020}. We can also rule out companions in the bright end of the red giant branch (RGB); for example, the ``tip'' of the RGB is at $\approx-7$ to $-6$ mag; \citealt{Valenti2004}). However, NIR-1 is consistent with giant stars slightly brighter than the RGB ``clump'', which has a $K$-band absolute magnitude range of $\approx-1$ to $-2$. NIR-1 is also consistent with the very luminous (massive) end of main sequence O stars, spanning $\approx -2.7$ to $-1$ mag \citep{PecautMamajek2013}; see Figure~\ref{fig:dists}.  However, given the $\sim 10$ Myr lifetime of such massive stars, and the likely nature of the FRB source as a neutron star (i.e., it had to have formed from an even more massive star to produce a compact object first), we consider such a scenario to be unlikely. The four sources at the edge of the $3\sigma$ localization region, spanning $\approx -5.1$ to $-3.4$ mag, are too luminous to be main sequence stars, but, like NIR-1, could be RGB stars.

\begin{figure*}[t!]
    \centering
    \includegraphics[width=\linewidth]{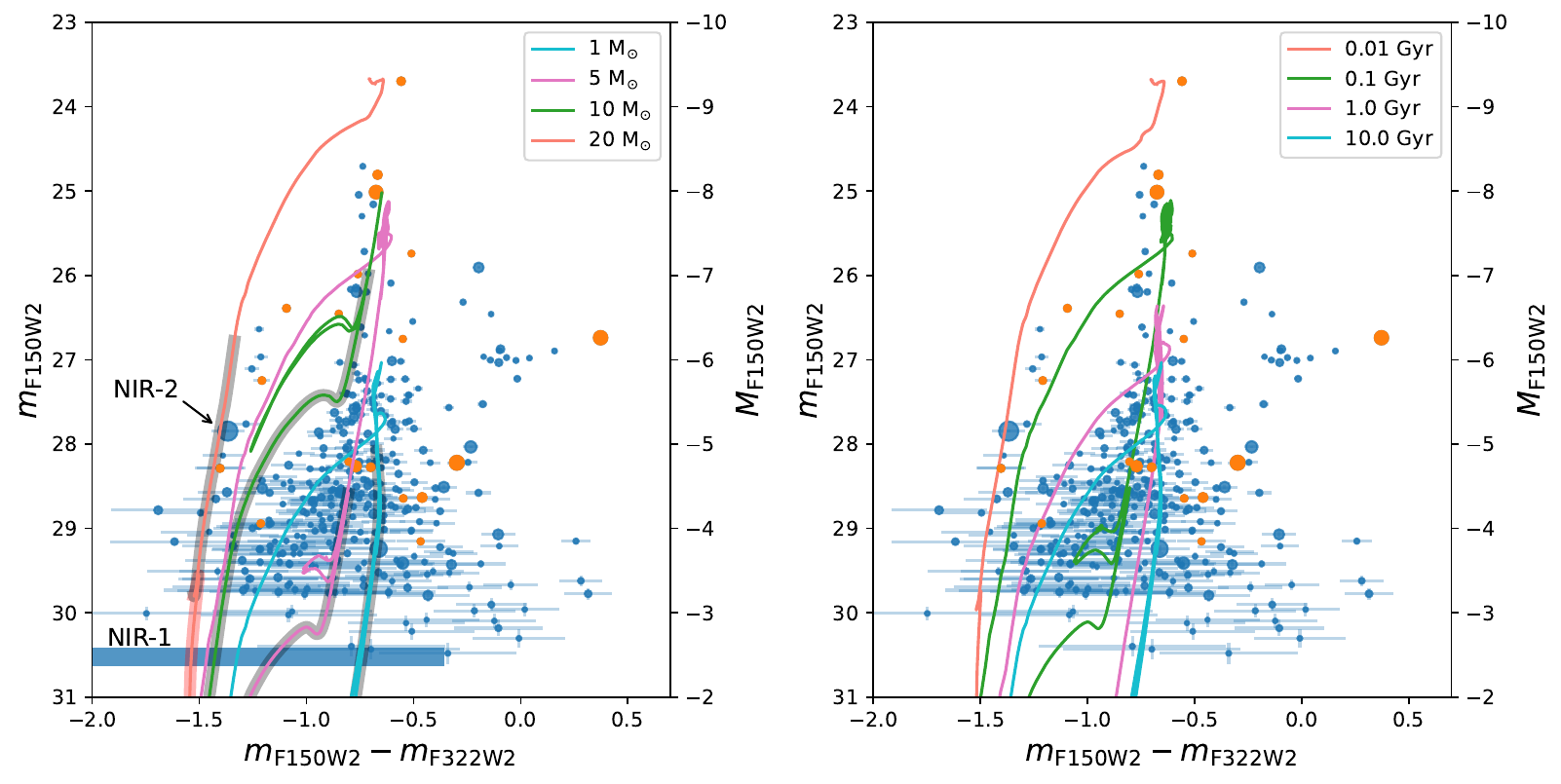}
    \caption{Color-magnitude diagram (CMD) of stellar sources within a radius of $3''$ ($570$ pc) of the location of \frb. The data points are sized by projected distance to \frb\ (larger for closer sources). Stars that appear spatially coincident with the nearby \ion{H}{2} region are shown as orange points. {\it Left:} Comparison to MIST evolutionary tracks for several initial masses: 1, 5, 10, 20 M$_\odot$. On each track, thick gray shading indicates the red giant branch; thick orange shading on the 20 M$_\odot$ track indicates the main sequence.  The location of NIR-1 on the CMD is shown as the horizontal blue bar, and is consistent with a $\sim 1-10$ M$_{\odot}$ red giant, $\sim 1-5$ M$_{\odot}$ post-AGB star, or a massive $\gtrsim 20$ M$_{\odot}$ main sequence star.  {\it Right:} Comparison to isochrones for several stellar population ages: 0.01, 0.1, 1, 10 Gyr. We find that the stellar population near the location of \frb\ contains both young ($\sim 10-100$ Myr) and old ($\sim 1-10$ Gyr) stars. If the actual FRB counterpart (e.g., isolated magnetar) remains undetected in our {\it JWST} data, it is possible that it originated in the younger population, which is dominated by stars in the nearby \ion{H}{2} region.  We note that NIR-2 is also consistent with this young population and may indicate that the young population overlaps the location of \frb.}
    \label{fig:CMD}
\end{figure*}

To further explore the possible evolutionary state and mass of a companion star with the luminosity and color of NIR-1, we compare it to evolutionary tracks from the MESA Isochrones and Stellar Tracks (MIST) repository \citep{Choi2016}, which provides synthetic photometry in the relevant {\it JWST} filters. In Figure~\ref{fig:CMD} we show stellar tracks (which include rotation) for stars at solar metallicity with initial masses of 1, 5, 10, and 20 M$_{\odot}$ (lower mass stars are fainter than our absolute magnitude limit at all evolutionary phases). For clarity, we do not show the short pre-main sequence phase, but otherwise show all phases that are brighter than our F150W2 magnitude limit.  Compared to these tracks, NIR-1 is consistent with a $\sim 1-10$ M$_{\odot}$ RGB star, $\sim 1-5$ M$_{\odot}$ post-asymptotic giant branch (AGB) star, or a massive $\gtrsim 20$ M$_{\odot}$ main sequence star.  The brighter sources on the edge of the $3\sigma$ localization region are also consistent with being RGB stars in a similar mass range, although at more evolved stages (or less advanced on the post-AGB phase).         

\subsection{Mass-Transferring Binary Models}
\label{sec:binary2}

The analysis in \S\ref{sec:binary} suggests that NIR-1 is consistent with either an evolved red giant or a very massive main sequence star. Binaries that produce RGB stars typically undergo unstable mass transfer, often leading to a common envelope (CE) phase. A possible channel for FRBs involves mass transfer in binary star systems involving a compact object \citep[e.g.,][]{Sridhar2022ApJ, Sridhar+24}. In such scenarios, accretion from the donor star may power flaring jets launched by the compact object. These unstable mass-transfer CE episodes can reach super-Eddington accretion rates, indicating that such binaries may represent a progenitor channel for FRBs.

\begin{figure}[t!]
    \centering
    \includegraphics[width=\linewidth]{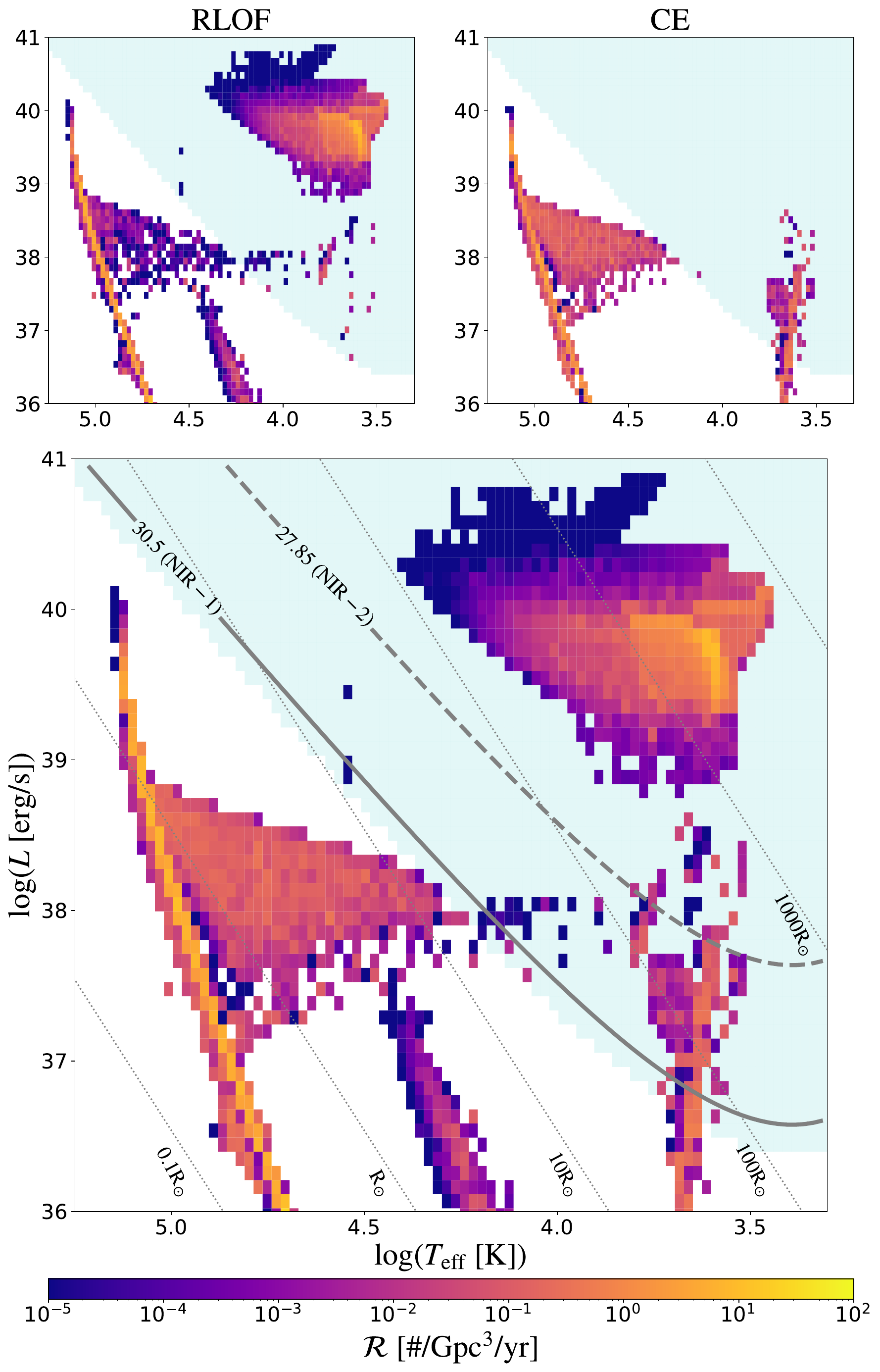}
    \caption{HR diagram of companion stars in binary systems with mass transfer onto a compact object, based on binary population synthesis models, color-coded by volumetric rate (Uno et al.~ in preparation). The main panel shows the entire population, while the top panels separate the population into stable mass transfer ({\it Left}; Roche lobe overflow: RLOF) and unstable mass-transfer ({\it Right}; common envelope: CE) systems.  In the main panel we highlight the locations of NIR-1 (solid contour) and NIR-2 (dashed contour); dotted lines show fixed blackbody radii. The blue shading marks the region ruled out as it leads to counterparts brighter than NIR-1.}
    \label{fig:HRmodels}
\end{figure}

To simulate a population of binary systems we employ the rapid population synthesis code COMPAS \citep{Riley2022ApJS}. Specifically, we use the dataset computed by \citet{vanSon+22}, which consists of $10^7$ binaries with zero-age main sequence (ZAMS) primary star masses of $10–150$ M$_{\odot}$, and metallicities from a logarithmically flat distribution of $10^{-4} \leq Z \leq 0.03$ (see also Uno et al., in preparation). In Figure~\ref{fig:HRmodels}, we show the resulting Hertzsprung-Russell (HR) diagram for the companion stars after the mass-transferring events in these binary models at their most luminous phase of evolution. 

Comparing this to the brightness of NIR-1 (or using it as an upper limit on a counterpart to \frb) we rule out supergiant companion (upper-right portion of the phase-space, mainly Roche Lobe overflow, RLOF, systems).  The systems consistent with the luminosity of NIR-1 (and the phase-space up to the luminosity of NIR-2) are mainly at the bright end of the RGB, with mass transfer via CE episodes, although some RLOF systems are possible as well; specifically, 84\% are helium Hertzsprung gap stars with binary ages of $\sim 30-50$ Myr. 

\subsubsection{Constraints on an Ultraluminous X-ray Source}

A subset of mass-transferring binaries may correspond to ultraluminous X-ray sources (ULXs), which are thought to be systems with high accretion rates.  \citet{Lopez2020} found NIR counterparts for 38 of 113 ULXs, all of which are brighter than $\approx -7$ mag in $H,K$-band, and thus much more luminous than NIR-1. These counterparts are consistent with RSGs, which we have ruled out for NIR-1. However, fainter NIR counterparts are not excluded for two-thirds of the ULX sample in \citet{Lopez2020} due to the limited depth of their search, and it is thus possible that NIR-1 may belong to such a fainter ULX population. 

We note that the X-ray follow-up of \frb\ about 11 d after the burst \citep{chandra} can also be used to place constraints on a ULX origin (see also \citealt{Cook25}). The X-ray flux upper limit is $F_{\rm X} \lesssim 7.6\times10^{-15}$ erg cm$^{-2}$ s$^{-1}$ ($0.5-10$ keV; 90\% confidence), corresponding to a luminosity limit of $L_{\rm X} \lesssim 1.5\times10^{39}$ erg s$^{-1}$, roughly at the limit of ULXs \citep{2022A&A...659A.188B}.

\subsection{Supernova Remnant}
\label{sec:snr}

We next explore the possibility that NIR-1 could be a supernova remnant or a pulsar wind nebula, associated with the compact object that produced \frb.  For example, SN\,1987A exhibits NIR continuum emission from its equatorial ring\footnote{The SN\,1987A equatorial ring, with an angular size of $\approx 2''$ would be unresolved at the distance of NGC\,4141.} with a peak flux density in the $JHK$ bands of $\approx 0.45$ mJy about 25 years after explosion (and about 25\% fainter at about 20 and 30 years; \citealt{kangas23}). This corresponds to an apparent brightness of $\approx 32$ AB mag in our F150W2 filter, about 4 times fainter than NIR-1. It is possible that a somewhat younger or more energetic SN than SN\,1987A would produce brighter NIR emission, in agreement with NIR-1. However, no SN has been discovered in NGC\,4141 at the position of \frb\ over the past $\sim 20$ years (see also \citealt{Cook25}), while two Type II SNe were discovered elsewhere in the galaxy in 2008 (SN\,2008X; \citealt{Boles2008}) and 2009 (SN\,2009E; \citealt{Boles2009}). We therefore consider this an unlikely origin for NIR-1. 

The Crab pulsar wind nebula (PWN) has an observed brightness of $\approx 2.5$ and $\approx 2.2$ mJy in $H$- and $K$-band, respectively \citep{ss09}, corresponding to an apparent brightness of $\approx 37$ AB mag in our F150W2 filter, $\approx 400$ times fainter than NIR-1.  We therefore consider it unlikely that NIR-1 is a PWN, unless it is significantly younger, but this again conflicts with the lack of a SN detection at the location of \frb\ over the past $\sim 20$ years.

It is, however, possible that if NIR-1 is unrelated to \frb, then an older SN remnant or PWN are present below the detection limit of our observations.

\subsection{Isolated Magnetar}
\label{sec:magnetar}

Magnetars have been invoked as possible emitting sources of FRBs. The NIR counterparts of Galactic magnetars span absolute magnitudes of $\approx 0-10$ (accounting for extinction; \citealt{2014ApJS..212....6O,2018ApJ...854..161L}), at least an order of magnitude less luminous than NIR-1.  This includes the variable NIR counterpart of SGR\,1935+2154, detected in 2015--2016 with an absolute magnitude of $\approx 8$ \citep{2018ApJ...854..161L}, or $\sim 10^{4}$ less luminous than NIR-1. We therefore consider it unlikely that NIR-1 represents direct emission from a magnetar, but we note that if NIR-1 is unrelated to the FRB, then our observations can accommodate an isolated magnetar as the origin of \frb.

\subsection{Properties of the Surrounding Stellar Population}
\label{sec:pop}

We next investigate the properties of the stellar population in the vicinity of the FRB position, in particular to shed light on its possible progenitor (e.g., an isolated magnetar) in the case that NIR-1 is a spurious association.  We construct a color-magnitude diagram (CMD) of sources within $3''$ ($570$ pc) of the location of \frb.  We select stellar (PSF-like) sources with a low level of crowding as determined by {\tt dolphot}. In Figure~\ref{fig:CMD} we show the resulting CMD compared to both MIST evolutionary tracks for stars of initial masses 1, 5, 10, and 20 M$_{\odot}$, as well as isochrones for stellar population ages of 0.01, 0.1, 1, and 10 Gyr. The isochrones and stellar tracks match well with the overall stellar population, accounting for the range of observed luminosities and colors.  In particular, we find that the stellar population consists of both young ($\sim 10-100$ Myr) and old ($\sim 1-10$ Gyr) stars.  The characteristic masses span from $\sim 1-20$ M$_{\odot}$, with the bulk of the stars residing on the RGB.

There is an \ion{H}{2} region centered $\approx 0.8''$ ($\approx 150$ pc) from the location of \frb\ (see Figure~\ref{fig:images}), also noted in ground-based imaging and spectroscopy \citep{Cook25}.  We roughly estimate the probability of chance coincidence for this \ion{H}{2} region by inspecting the number of similar regions in an annulus of inner and outer radii of $\approx 20$ and $40''$ from the center of NGC\,4141 (which encompases the location of \frb).  We find a number of $\approx 20$ such regions per $\approx 10^3$ arcsec$^{2}$, leading to $P_{\rm cc}\approx 0.05$ in relation to the FRB position (see \S\ref{sec:id}), which may suggest an association with \frb. 

The \ion{H}{2} region is well resolved in our images, exhibiting both isolated stars and diffuse emission, and with a rough size of its central region of $\approx 120$ pc (Figure~\ref{fig:images}). In Figure~\ref{fig:CMD} we highlight on the CMD stars that are spatially coincident with this \ion{H}{2} region. These stars account for nearly half of the most luminous stars within the 570-pc region around \frb.  Overall, they predominantly span the younger end of the stellar population ages, $\lesssim 0.1$ Gyr, and appear to be consistent mostly with masses of $\gtrsim 5$ M$_\odot$ (Figure~\ref{fig:CMD}). Indeed, the brightest source in the \ion{H}{2} region is likely a $\sim 20$ M$_\odot$ supergiant with an age of $\sim 10$ Myr. This suggests that if the progenitor of the compact object that produced \frb\ was born in the same stellar population, then its initial mass was $\gtrsim 20$ M$_\odot$.

We note that the brightest of the four sources at the edge of the $3\sigma$ localization region (NIR-2) is also consistent with a massive star in this young population based on its location in the CMD (Figure~\ref{fig:CMD}) suggesting this population extends over an area of $\gtrsim 100$ pc beyond the nominal ``core'' of the \ion{H}{2} region (dominated by the brightest stars).  Therefore, it is plausible that the FRB-producing compact object was born at or near its current location and that it experienced little to no natal kick.

\subsection{Afterglow}
\label{sec:ag}

In scenarios where FRBs are accompanied by a significant mass ejection, (e.g., from a magnetar flare; \citealt{Beloborodov17,Metzger+19}), the flare ejecta can produce a synchrotron ``afterglow'' as it interacts with an external medium surrounding the magnetar \citep{Metzger+19}.  Given the FRB radio energy of $E_{\rm FRB}\approx 10^{39}$ erg \citep{Cook25}, we infer a total outburst energy of $E = E_{\rm FRB}/f_{r}\sim 10^{44}$ erg, for a radio efficiency of $f_{r}= 10^{-5}$, matching the value inferred from the X-ray counterpart of the Galactic SGR 1935+2154 burst (e.g., \citealt{Scholz2020ApJ...901..165S, Margalit2020ApJ...899L..27M}).  We assume that the initially relativistic ejecta expands into a homogeneous external medium of number density $n_{\rm ext} = n_{\rm 0}$ cm$^{-3}$, causing the ejecta to decelerate and become mildly relativistic at a radius (e.g., \citealt{Nakar&Piran11}),
\be
R_{\rm dec} \approx \left(\frac{3E}{4\pi n_{\rm ext}m_p c^{2}}\right)^{1/3} \approx 2.5\times 10^{15}\,{\rm cm} \,E_{44}^{1/3}n_{0}^{-1/3}
\ee
on a timescale,
\be
t_{\rm dec} \approx \frac{R_{\rm dec}}{c} \approx 9\times 10^{4}\,{\rm s}\,E_{44}^{1/3}n_{0}^{-1/3},
\ee
i.e., within a few hours of the burst. At times $t\gg t_{\rm dec}$, relevant to our observations, the blastwave is in the Sedov-Taylor phase. Electrons accelerated into a power-law distribution of velocities at the shock will emit synchrotron radiation with a characteristic frequency that decreases with time (e.g., \citealt{Metzger+15}),
\be
\nu_{\rm m} \approx 0.03\,{\rm GHz} \,E_{44}^{1/2}\epsilon_{\rm e,-1}^{2}\epsilon_{\rm B,-1}^{1/2}\left(\frac{t}{2\,{\rm month}}\right)^{-3/2},
\ee
where $\epsilon_{\rm e} = 0.1\epsilon_{\rm e,-1}$ and $\epsilon_{\rm B} = 0.1\epsilon_{\rm B,-1}$ are the fraction of the shock's energy in relativistic electrons and magnetic fields, respectively. At NIR frequencies, $\nu_{\rm obs} \gg \nu_{\rm m}$, the specific flux evolves as (e.g., \citealt{Nakar&Piran11}):
\begin{multline*}
\nu L_{\nu} \approx 2.6\times 10^{34}\,{\rm erg\,s^{-1}}\,E_{44} n_0^{0.8}\epsilon_{\rm B,-1}^{0.8}\epsilon_{\rm e,-1}^{1.3}\nu_{\rm obs,14}^{0.35} \\
\times\left(\frac{t}{t_{\rm dec}}\right)^{-1.35} 
\end{multline*}
\begin{multline}
\approx  1.1\times 10^{32}\,{\rm erg\,s^{-1}}\,E_{44}^{1.45} n_0^{0.35}\epsilon_{\rm B,-1}^{0.8}\epsilon_{\rm e,-1}^{1.3}\nu_{\rm obs,14}^{0.35} \\
\times\left(\frac{t}{2\,{\rm month}}\right)^{-1.35},
\end{multline}
where we assume a typical value of $p = 2.3$ for the power-law index of the electron energy distribution ($dN/d\gamma_{\rm e} \propto \gamma_{\rm e}^{-p}$).  This is an optimistic estimate, as it neglects the potential effects of a cooling break between the radio and NIR bands. Even so, this luminosity is already orders of magnitude lower than NIR-1 ($\nu L_\nu\approx 4.5\times 10^{35}$ erg s$^{-1}$), and therefore cannot explain its origin, unless the outburst energy is $\sim 300$ times larger than our fiducial value (i.e., $f_r\sim 10^{-8}$).  We therefore consider an afterglow origin to be unlikely.   

\subsection{Dust Echo}
\label{sec:dust}

An alternative possibility is that the afterglow's optical, UV, or X-ray radiation is reprocessed by nearby dust into emission in the NIR band. Assuming silicate dust grains of size $a = 1a_{\mu m}$ $\mu m$ at radius $r$, the resulting dust temperature is (e.g., \citealt{Metzger&Perley23}):
\begin{eqnarray}
T_{\rm d} &\approx& \left(\frac{L}{16\pi \sigma \langle Q_{\rm IR}\rangle _{\rm T} r^{2}}\right)^{1/4} \nonumber \\
&\approx& 3.1\times 10^3{\rm K}\,L_{43}^{1/5}a_{\mu m}^{-1/5}r_{16}^{-2/5},
\end{eqnarray}
where $\langle Q_{\rm IR}\rangle_{\rm T}$ is the IR emission efficiency (e.g., \citealt{Draine&Lee84}), and we have assumed a duration of 10 s for the emission (i.e., $10^{43}$ erg s$^{-1}$). Equating $T_d$ to the sublimation temperature, $T_{\rm s} = 1700\, T_{\rm s,1700}$ K, we find that dust is sublimated interior to the radius,
$r_{\rm s} \simeq 4.4\times 10^{16}\,{\rm cm}\,L_{43}^{1/2}a_{\mu m}^{-1/2}T_{\rm s,1700}^{-5/2}$. At $r > r_{\rm s}$, the heated dust temperature distribution follows
\be
T_{\rm d} = T_{\rm s}\left(\frac{r}{r_{\rm s}}\right)^{-2/5}.
\ee
Dust of temperature $T_{\rm d}$ re-emits the absorbed light at IR frequencies, $h \nu \approx 4 kT_{\rm d}$.  This results in a rough mapping between IR wavelength and emission radius:
\be
\lambda \approx \frac{h c}{4k T_{\rm D}} \approx 2.1\mu m \, \left(\frac{r}{r_{\rm s}}\right)^{2/5}.
\ee
The radiation reprocessed by a spherical dust shell of radius $r$ is received by an observer over a characteristic time interval $t\approx 2r/c$ (e.g., \citealt{Lu+16,vanVelzen+16}), resulting in a wavelength-time mapping:
\begin{eqnarray}
\lambda &\approx& 2.1\mu m \, \left(\frac{t}{2r_{\rm s}/c}\right)^{2/5} \nonumber \\
&\simeq& 2.7\mu m\left(\frac{t}{2\,{\rm month}}\right)^{2/5}L_{43}^{-1/5}a_{\mu m}^{1/5}T_{\rm s,1700}.
\end{eqnarray}

Thus, we would expect emission in the bandpass of our {\it JWST} observations at $t = 2$ months for a wide range of transient luminosities, $L\sim (10^{41}-10^{44})a_{\mu m}$ erg s$^{-1}$.

If a fraction $f_{\rm abs}$ of the transient's radiated energy is absorbed by the dust at radius $r = ct$, the NIR echo luminosity at $t\approx 2$ months, is given by:
\begin{equation}
L_{\rm IR} \approx \frac{f_{\rm abs}E}{2t} \approx 10^{36}\,{\rm erg\,s^{-1}}f_{\rm abs,-1}E_{44}
\end{equation}
Thus, for our fiducial assumptions, it is plausible to produce a luminosity comparable to that of NIR-1 ($\approx 4.5\times 10^{35}$ erg s$^{-1}$). This is of course, provided that the assumption of an accompanying optical/UV/X-ray outburst is correct and that there is sufficient dust surrounding the source at the right scale ($r \sim ct/2 \sim 10^{17}$ cm) to absorb a sufficient fraction ($f_{\rm abs}\sim 0.1$) of the outburst emission. A clear prediction of this model is that NIR-1 will fade away in a repeated observation $\gtrsim {\rm year}$ later.

\section{Summary and Conclusions}
\label{sec:conc}

We presented deep {\it JWST} observations of the localization region of \frb\ in NGC\,4141 to search for a NIR counterpart. The combination of exquisite localization precision ($\lesssim 0.1''$), relative proximity ($\approx 40$ Mpc), and unprecedented depth and angular resolution with {\it JWST} produce some of the most compelling insights on the nature of the FRB system.  We identified a source (NIR-1) located $\approx40$ mas from the localization centroid of \frb, which is the only source within $\approx2.7\sigma$ of the localization centroid.  The nearest sources other than NIR-1 are on the edge of the $3\sigma$ confidence region, and while brighter than NIR-1, they do not substantially alter our conclusions.

Based on the properties of NIR-1 (or using its brightness as an upper bound), we rule out the possibility of a globular cluster, young star cluster, red supergiant star, red giant at or below the tip of the RGB, an isolated magnetar, or a supernova remnant / pulsar wind nebula.  We find that NIR-1 is consistent with being a red giant star near the RGB ``clump'', or a $\gtrsim 20$ M$_{\odot}$ main sequence star, although we note that a massive main sequence star is less likely given the short lifetimes of these stars.  We further explore the stellar population in a region of about 570 pc radius around the location of \frb, including the resolved stellar population of an \ion{H}{2} region centered $\approx 150$ pc from the localization centroid. Based on this information, we conclude with the following three possibilities for the progenitor of \frb:

\begin{itemize}
    \item NIR-1 is associated with \frb, and represents the evolved giant star (possibly in a common envelope mass transfer phase) or very massive main sequence binary companion of an FRB-producing compact object. The other four sources on the outskirts of the $3\sigma$ localization region are also consistent with being evolved giants.

    \item NIR-1 coincides with the localization region by chance, and a counterpart of \frb\ was not detected, leaving open the possibility of a less luminous binary companion, or an isolated neutron star / magnetar.  In this scenario, the proximity to an \ion{H}{2} region (with $P_{\rm cc}\approx 0.05$) that contains young ($\sim 10-100$ Myr) massive stars (up to $\sim 20$ M$_\odot$) would point to a compact object engine produced in the core-collapse of a massive star. The location of young stars in the region of \frb\ (e.g., NIR-2) indicate that the compact object could have formed {\it in-situ} with no natal kick.

    \item NIR-1 represents a dust echo from an energetic outburst that accompanied \frb\ (e.g., mass ejection from a magnetar flare).  This requires sufficient dust at a scale of $\sim 0.03$ pc and an outburst of $\sim 10^{44}$ erg.  This hypothesis can be tested with follow-up {\it JWST} observations on a timescale of $\gtrsim {\rm year}$.    
\end{itemize}

Our {\it JWST} study of \frb\ demonstrates the power of FRB follow-up observations at a scale of $\sim 10$ pc, as well as associated studies of the resolved stellar populations at the locations of FRBs.  Future studies with {\it JWST} would benefit from a larger sample of events, as well as more rapid and multi-epoch follow-up to potentially capture and characterize transient emission.

\begin{acknowledgements}
The Berger Time-Domain Group at Harvard is supported by NSF and NASA grants.  This work is based on observations made with the NASA/ESA/CSA {\it James Webb Space Telescope}. The data were obtained from the Mikulski Archive for Space Telescopes at the Space Telescope Science Institute, which is operated by the Association of Universities for Research in Astronomy, Inc., under NASA contract NAS 5-03127 for {\it JWST}. These observations are associated with program \#9331.  C.D.K.~gratefully acknowledges support from the NSF through AST-2432037, the HST Guest Observer Program through HST-SNAP-17070 and HST-GO-17706, and from JWST Archival Research through JWST-AR-6241 and JWST-AR-5441.  B.D.M.~acknowledges support from the NSF (grant AST-2406637) and the Simons Foundation (grant 727700).  N.S.~is supported by a grant from the Simons Foundation (MP-SCMPS-00001470).  A.M.C.~is a Banting Postdoctoral Researcher.  Y.D.~is supported by the National Science Foundation (NSF) Graduate Research Fellowship under grant No.~DGE-2234667.  W.F.~gratefully acknowledges support by the National Science Foundation under grant Nos. AST-2206494, AST-2308182, and CAREER grant No. AST-2047919, the David and Lucile Packard Foundation, the Alfred P. Sloan Foundation, and the Research Corporation for Science Advancement through Cottrell Scholar Award \#28284.  V.M.K.~holds the Lorne Trottier Chair in Astrophysics \& Cosmology, a Distinguished James McGill Professorship, and receives support from an NSERC Discovery grant (RGPIN 228738-13).  C. L.~acknowledges support from the Miller Institute for Basic Research at UC Berkeley.  K.W.M.~holds the Adam J.~Burgasser Chair in Astrophysics and received support from an NSF grant (2018490).  K.N.~is an MIT Kavli Fellow.  A.B.P.~is a Banting Fellow, a McGill Space Institute~(MSI) Fellow, and a Fonds de Recherche du Quebec -- Nature et Technologies~(FRQNT) postdoctoral fellow.  V.S.~is supported by a Fonds de Recherche du Quebec—Nature et Technologies (FRQNT) Doctoral Research Award.  K.S.~is supported by the NSF Graduate Research Fellowship Program.  S.S.~gratefully acknowledges the support by the Brinson Foundation as a joint NU-UC Brinson Postdoctoral Fellow.
\end{acknowledgements}

\software{Astropy \citep{astropy}, {\tt dolphot} \citep{dolphot,JWST-dolphot}, {\tt JHAT} \citep{JHAT}, {\tt WebbPSF} \citep{STPSF}}

\facilities{{\it JWST} (NIRCam)}

\bibliography{references}
\bibliographystyle{aasjournalv7}

\end{document}